\documentclass[12pt] {iopart}
\begin{document}

\title[]{Periodic and Localized Solutions of the Long Wave-Short Wave Resonance Interaction
Equation}

\author{R. Radha$^{1}$,   C. Senthil  Kumar$^{2}$, M.
Lakshmanan$^{2}$, X. Y. Tang$^{3}$ and S. Y.
Lou$^{3,4}$}
\address{$^1$  Dept. of Physics, Govt. College for Women, Kumbakonam - 612 001, India}
\address{$^2$  Centre for Nonlinear Dynamics, Dept. of
Physics, Bharathidasan University,Tiruchirapalli - 620 024, India.}
\address{$^3$  Dept. of Physics, Shangai Jiao Tong University, Shanghai 200030, China}
\address{$^4$  Center of Nonlinear Science, Ningbo
University, Ningbo 315211, China} 

\ead{lakshman@cnld.bdu.ac.in}

\begin{abstract}
In this paper, we investigate the (2+1) dimensional long wave-short wave
resonance interaction (LSRI) equation and show that it possess the
Painlev\'e property.  We then solve the LSRI equation using Painlev\'e
truncation approach through which we are able to construct solution in 
terms of three arbitrary functions. Utilizing the arbitrary functions
present in the solution, we have generated a wide class of elliptic function periodic wave solutions and
exponentially localized solutions such as dromions, multidromions, instantons, multi-instantons
and bounded solitary wave solutions.
\end{abstract}

\maketitle

\section{Introduction}
The identification of dromions which are exponentially decaying solutions in the
Davey-Stewartson I
and other equations [1-6] has triggered a renewed interest in the study of
integrable models in (2+1) dimensions.  Dromions arise essentially by virtue of
coupling the field variable to a mean field/potential, thereby preventing wave
collapse in (2+1) dimensions and they can in general undergo inelastic collision
unlike one dimensional solitons.  The identification of a large number of
arbitrary functions in the solutions of (2+1) dimensional integrable models has only added to the
richness in the structure of them and hence construction of localized
excitations in (2+1) dimensions continues to be a challenging and rewarding
contemporary problem.

In this paper, we consider the existence of localized structures in the 
long wave short wave resonance interaction equation of the form
\numparts
\begin{eqnarray}
i(S_{t}+S_{y})-S_{xx}+LS=0, \\ 
L_{t}=(2SS^*)_{x}, 
\end{eqnarray} \label{I} 
\endnumparts
where the fields $S$ and $L$  denote short surface wave packets and
long interfacial waves respectively, while $*$ stands for complex conjugation. The above
equation has been recently studied [7,8] and its positon and one dromion
solutions have been generated through Hirota method.  However, no further general solutions could be constructed through this procedure for equations (\ref{I}). In this
contribution, we develop a very simple and straightforward procedure to generate a rather extended class of generic solutions of physical interest.   For this purpose, first we carry out the singularity structure analysis to the
LSRI equation and confirm its Painlev\'e nature.  We utilize the local
Laurent expansion of the general solution and truncate
it at the constant level term (Painlev\'e truncation approach) and
obtain solutions in terms of arbitrary functions. Through this procedure
we generate various periodic and exponentially localized solutions to equation (\ref{I}).  The novelty here is that 
the solution is generated through a very simple procedure but the solution
obtained is rich in structure because of the arbitrary functions [9-14] present
in the solution.

The plan of the paper is as follows.  In Section 2, we present the singularity
structure analysis of the LSRI equation.  Using these results, in Section 3, we have shown the  
construction of solutions for the LSRI equation through Painlev\'e truncation approach.
Section 4 contains a wide class of localized solutions of the LSRI equation both periodic and 
exponentially localized ones, through judicial choice of the arbitrary functions.  In Section 5, 
we summarize our results.  The appendix contains the one dromion solution of LSRI equation obtained
through Hirota bilinearization approach for comparison.

\section{Singularity Structure Analysis}

To explore the singularity structure of equation (\ref{I}), we rewrite
$S$=q and $S^*$=r to obtain the following set of coupled equations,
\numparts
\begin{eqnarray}
i(q_{t}+q_{y})-q_{xx}+Lq=0, \\ 
-i(r_{t}+r_{y})-r_{xx}+Lr=0,  \\ 
L_{t}=(2qr)_{x}. 
\end{eqnarray} \label{s1} 
\endnumparts
We now effect a local Laurent expansion in the neighbourhood of a
noncharacteristic singular manifold $\phi(x,y,t)$=0,
$\phi_{x}\neq0$, $\phi_{y}\neq0$. Assuming the leading orders of the solutions
of equation (\ref{s1}) to have the form
\begin{equation}
q=q_{0}\phi^{\alpha},r=r_{0}\phi^{\beta},L=L_{0}\phi^{\gamma}, \label{s2} 
\end{equation}
where $q_{0}$, $r_{0}$ and $L_{0}$ are analytic functions of ($x$, $y$, $t$) and
$\alpha$, $\beta$, $\gamma$ are integers to be determined, we substitute
(\ref{s2}) into (\ref{s1}) and balance the most dominant terms to obtain
\begin{equation}
\alpha=\beta=-1,\gamma=-2,
\end{equation}
with the condition
\begin{equation}
q_{0}r_{0}=\phi_{x}\phi_{t}, L_{0}=2 \phi_{x}^2. \label{s6}
\end{equation}
Now, considering the generalized Laurent expansion of the solutions in the
neighbourhood of the singular manifold,
\numparts
\begin{eqnarray}
q=q_{0}\phi^{\alpha}+...+q_{j}\phi^{r+\alpha}+..., \\
r=r_{0}\phi^{\beta}+...+r_{j}\phi^{r+\beta}+..., \\
L=L_{0}\phi^{\gamma}+...+L_{j}\phi^{r+\gamma}+..., 
\end{eqnarray} \label{eq6}
\endnumparts
the resonances which are the powers at which arbitrary functions
enter into (6) can be determined by substituting (6) into (\ref{s1}).
Vanishing of the coefficients of
($\phi^{j-3}$,$\phi^{j-3}$,$\phi^{j-3}$) lead to the condition
\begin{equation}
\left(
\begin{array}{ccc}
-j(j-3)\phi_{x}^2 & 0 & q_{0} \\
0 & -j(j-3)\phi_{x}^2 & r_{0} \\
2(j-2)r_{0}\phi_{x} & 2(j-2)q_{0}\phi_{x} & -(j-2)\phi_{x}
\end{array}
\right)
\left(
\begin{array}{c}
q_{j} \\
r_{j} \\
L_{j}
\end{array}
\right) = 0. \label{s4}
\end{equation} 
From equation (\ref{s4}), one gets the resonance values as
\begin{equation}
j=-1,\ 0,\ 2,\ 3,\ 4. 
\end{equation}
The resonance at $j$ = -1 naturally represents the arbitrariness
of the manifold $\phi(x,y,t)=0$. In order to prove the existence
of arbitrary functions at the other resonance values, we now
substitute the full Laurent series
\numparts
\begin{eqnarray}
q=q_{0}\phi^{\alpha}+\sum_{j}q_{j}\phi^{j+\alpha}, \\
b=r_{0}\phi^{\beta}+\sum_{j}r_{j}\phi^{j+\beta}, \\
L=L_{0}\phi^{\gamma}+\sum_{j}L_{j}\phi^{j+\gamma} 
\end{eqnarray} \label{s13}
\endnumparts
into equation (\ref{s1}).  Now collecting the coefficients of
($\phi^{-3}$, $\phi^{-3}$, $\phi^{-3}$) and solving the resultant
equation, we obtain equation (\ref{s6}), implying the existence of a
resonance at $j=0$.

Similarly collecting the  coefficients of ($\phi^{-2}$,
$\phi^{-2}$, $\phi^{-2}$) and solving the resultant equations by
using the Kruskal's ansatz, $\phi(x,y,t)=x+\psi(y,t)$, we get
\numparts
\begin{eqnarray}
q_{1}=\frac{1}{2}[iq_{0}(\psi_{t}+\psi_{y})-2q_{0x}],\\
r_{1}=\frac{1}{2}[-ir_{0}(\psi_{t}+\psi_{y})-2r_{0x}], \label{s14} \\
L_{1}=0. 
\end{eqnarray} 
\endnumparts

Collecting the coefficients  of ($\phi^{-1}$,$\phi^{-1}$,$\phi^{-1}$), we have
\numparts
\begin{eqnarray}
i(q_{0t}+q_{0y})-q_{0xx}+L_{0}q_{2}+L_{1}q_{1}+L_{2}q_{0}=0,\label{s7} \\
-i(r_{0t}+r_{0y})-r_{0xx}+L_{0}r_{2}+L_{1}r_{1}+L_{2}r_{0}=0,\label{s8} \\
L_{1t}=2[q_{0x}r_{1}+r_{0x}q_{1}+q_{1x}r_{0}+q_{1}r_{0x}]=0.
\end{eqnarray} 
From (\ref{s7}) and (\ref{s8}), we can eliminate $L_{2}$ to obtain a single equation for
the two unknowns $q_2$ and $r_2$,
\begin{eqnarray}
L_{0}(r_{0}q_{2}-q_{0}r_{2})-(r_{0}q_{0xx}-q_{0}r_{0xx})
\nonumber \\
\;\;\;\;\;\;\;\;\;\;\; +i(r_{0}(q_{0t}+q_{0y}) +q_{0}(r_{0t}+r_{0y}))=0   
\end{eqnarray} 
\endnumparts
which ensures that either $q_{2}$ or $r_{2}$ is arbitrary.  Obviously $L_2$
itself can be obtained either from (\ref{s7}) or (\ref{s8}).
Similarly, collecting the coefficients  of ($\phi^{0}$,$\phi^{0}$,$\phi^{0}$), we
have
\numparts
\begin{eqnarray}
i(q_{1t}+q_{2}\psi_{t})+i(q_{1y}+q_{2}\psi_{y})-(q_{1xx}+2q_{2x})
+L_{2}q_{1}+L_{3}q_{0}=0,\label{s9} \\
-i(r_{1t}+r_{2}\psi_{t})-i(r_{1y}+r_{2}\psi_{y})-(r_{1xx}+2r_{2x})
+L_{2}r_{1}+L_{3}r_{0}=0,\label{s10} \\
L_{2t}+L_{3}\psi_{t}=2[q_{0x}r_{2}+(q_{1x}+q_{2})r_{1}+(q_{2x}+2q_{3})r_{0}
\nonumber\\
\;\;\;\;\;\;\;\;\;\;\;\;\;\;\;\;\;\;\;\;\;\;\;\; +r_{0x}q_{2}+(r_{1x}+r_{2})q_{1}+(r_{2x}+2r_{3})q_{0}].
\end{eqnarray}  
We rewrite the  equations (\ref{s9}) and (\ref{s10}) as
\begin{eqnarray}
L_3=\frac{1}{q_0}(-i(q_{1t}+q_{2}\psi_{t})-i(q_{1y}+q_{2}\psi_{y})+(q_{1xx}+2q_{2x})
-L_{2}q_{1}), \label{s11} \\
L_3=\frac{1}{r_0}(+i(r_{1t}+r_{2}\psi_{t})+i(r_{1y}+r_{2}\psi_{y})+(r_{1xx}+2r_{2x})
-L_{2}r_{1}). \label{s12} 
\end{eqnarray} 
\endnumparts
Making use of the earlier relations (\ref{s6}), (\ref{s14}) and (\ref{s8}), we find
that the right hand sides of equations (\ref{s11}) and (\ref{s12}) are equal. 
Then, we are left with two equations for three unknowns.  So, one of the three 
coefficients $q_{3}$, $r_{3}$ or $L_{3}$ is arbitrary. Collecting now the 
coefficients  of ($\phi$, $\phi$, $\phi$), we have
\numparts
\begin{eqnarray}
i(q_{2t}+2q_{3}\psi_{t})+i(q_{2y}+2q_{3}\psi_{y})-(q_{2xx}+4q_{3x}+6q_{4})
\nonumber \\
\;\;\;\;\;\;\;\;\;\;\; +L_{0}q_{4}+L_{2}q_{2}+L_{3}q_{1}+L_{4}q_{0}=0, \\
-i(r_{2t}+2r_{3}\psi_{t})-i(r_{2y}+2r_{3}\psi_{y})-(r_{2xx}+4r_{3x}+6r_{4})
\nonumber \\
\;\;\;\;\;\;\;\;\;\;\; +L_{0}r_{4}+L_{2}r_{2}+L_{3}r_{1}+L_{4}r_{0}=0, \\
L_{3t}+2L_{4}\psi_{t}=2[q_{0x}r_{3}-q_{0}r_{4}+(q_{1x}+q_{2})r_{2} \nonumber \\
\;\;\;\;\;\;\;\;\;\;\;\;\;\;\;\;\;\;\;\;\;\;\;\; +(q_{2x}+2q_{3})r_{1}+(q_{3x}+3q_{4})r_{0}+r_{0x}q_{3}-r_{0}q_{4} \nonumber \\
\;\;\;\;\;\;\;\;\;\;\;\;\;\;\;\;\;\;\;\;\;\;\;\; +(r_{1x}+r_{2})q_{2}+(r_{2x}+2r_{3})q_{1}+(r_{3x}+3r_{4})q_{0}]. 
\end{eqnarray} 
\endnumparts
Here also, the above set of three equations reduces to two equations.
So, one of the three functions $q_{4}$, $r_{4}$ or $L_{4}$ is arbitrary.
One can proceed further to determine all other coefficients of the Laurent
expansions (9) without the introduction of any movable critical singular manifold.
Thus, the LSRI equation indeed satisfies the Painlev\'e property. 

\section{Painlev\'e Truncation approach}
To generate the solutions of LSRI equation, we suitably harness the results of
the Painlev\'e analysis.  Truncating the Laurent series of the solutions of the
LSRI equation at the constant level term, we have the B\"acklund transformation
\numparts
\begin{eqnarray}
q=\frac{q_0}{\phi}+q_1,\\
r=\frac{r_0}{\phi}+r_1,\\
L=\frac{L_0}{\phi^2}+\frac{L_1}{\phi}+L_2.
\end{eqnarray} \label{p1}
\endnumparts
Assuming a seed solution given by
\begin{equation}
q_1=r_1=0, \quad L_2=L_2(x,y), \label{p2}
\end{equation}
we now substitute (14) with the above seed solution (\ref{p2}) 
into equations (\ref{s1}) to obtain the following system of equations 
by equating the coefficients of $(\phi^{-3},\phi^{-3},\phi^{-3})$,
\numparts
\begin{eqnarray}
-2q_0 \phi_x^2+L_0q_0 =0,\\
-2r_0 \phi_x^2+L_0r_0 =0,\\
L_0 \phi_t = 2 q_0 r_0 \phi_x.
\end{eqnarray} 
\endnumparts
Solving the above system of equations, we obtain the leading order coefficients
already given by equation (\ref{s6}), namely $q_{0}r_{0}=\phi_{x}\phi_{t}$ and  
$L_{0}=2 \phi_{x}^2$. Now collecting the coefficients
$(\phi^{-2},\phi^{-2},\phi^{-2})$
we have the following system of equations,
\numparts
\begin{eqnarray}
-iq_0\phi_t-iq_0\phi_y+2q_{0x}\phi_x+q_0\phi_{xx}+L_1q_0=0, \label{p5} \\
ir_0\phi_t+ir_0\phi_y+2r_{0x}\phi_x+r_0\phi_{xx}+L_1r_0=0,\label{p6} \\
L_{0t}-L_1 \phi_t = 2 (q_0r_0)_x. \label{p3}
\end{eqnarray} \label{p7}
\endnumparts
From equation (\ref{p3}), we have
\begin{equation}
L_1=-2\bigg[\phi_{xx}+\frac{\phi_x \phi_{tx}}{\phi_t}\bigg]  \label{p4}
\end{equation}
Using (\ref{p4}) in equation (\ref{p5}) or (\ref{p6}), one can easily obtain the relation
\begin{equation}
\frac{q_{0x}}{q_0}=\frac{1}{2}\bigg[\frac{i(\phi_t+\phi_y)+\phi_{xx}
-\frac{2\phi_x\phi_{tx}}{\phi_t}}{\phi_x}\bigg]
\end{equation}
On integration, we obtain
\begin{equation}
q_0=F(y,t)\mbox{exp}\bigg[{\frac{1}{2}\int\frac{i(\phi_t+\phi_y)+\phi_{xx}
-\frac{2\phi_x\phi_{tx}}{\phi_t}}
{\phi_{x}}{\rm d}x}\bigg], \label{p4_1}
\end{equation}
where $F(y,t)$ is an arbitrary function of $y$ and $t$.
Obviously the above solution is consistent with (17).  

Again collecting the coefficients of $(\phi^{-1},\phi^{-1},\phi^{-1})$, we have the following set of
equations
\numparts
\begin{eqnarray}
iq_{0t}+iq_{0y}-q_{0xx}+L_2q_0 =0, \label{p13} \\
-ir_{0t}-ir_{0y}-r_{0xx}+L_2r_0 =0, \label{p14} \\
L_{1t}=0. \label{p8}
\end{eqnarray} 
\endnumparts
Using (\ref{p4}), we rewrite equation (\ref{p8}) to obtain the trilinear form
\begin{equation}
\phi_t^2\phi_{xxt}-\phi_x\phi_{tx}\phi_{tt}+\phi_{xt}^2\phi_t
+\phi_{x}\phi_{ttx}\phi_{t}=0. \label{p9}
\end{equation}
The structure of the trilinear equation (\ref{p9}) suggests that
a specific solution can be given in the form
\begin{equation}
\phi = \phi_1(x,y)+\phi_2 (y,t), \label{p10}
\end{equation} 
where $\phi_1(x,y)$ and $\phi_2 (y,t)$ are arbitrary functions in the indicated 
variables. Using (\ref{p10}) in equations (\ref{p4}) and (\ref{p4_1}), one can 
obtain the functions $q_0$ and $L_1$ as
\numparts
\begin{eqnarray}
q_0=F(y,t)\mbox{exp}\bigg[{\frac{1}{2}\int\frac{i(\phi_{2t}+\phi_{1y}+\phi_{2y})+\phi_{1xx}}
{\phi_{1x}}{\rm d}x}\bigg], \label{p12} \\
L_1=-2\phi_{1xx}. \label{p11}
\end{eqnarray} 
\endnumparts
From (\ref{p11}), we find that equation (\ref{p8})  is an identity.  
Using (\ref{p12}), equations (\ref{p13}) and (\ref{p14})
can be reduced to the form
\begin{equation}
\phi_{2tt}+\phi_{2ty}=0. \label{p11_1}
\end{equation}
Equation (\ref{p11_1}) can be solved readily to express the submanifold
$\phi_2(y,t)$ in the form
\begin{equation}
\phi_2 = F_2(y)+F_3(t-y), \label{p16}
\end{equation}
where $F_2(y)$ and $F_3(t-y)$ are arbitrary functions in $y$ and $(t-y)$,
respectively.

Finally, collecting the coefficients of $(\phi^{0},\phi^{0},\phi^{0})$,
we have only one equation
\begin{equation}
L_{2t}=0. \label{p15}
\end{equation}
Using (\ref{p13}) for $L_2$, (\ref{p15}) reduces to the form
\begin{equation}
(F_{tt}+F_{ty})F+(F_t+F_y) F_t =0. \label{p15_1}
\end{equation} 
Equation (\ref{p15_1}) can be solved to obtain the form for $F(y,t)$ as
\begin{equation}
F(y,t)=F_1(t-y). 
\end{equation}
Thus the LSRI equation (\ref{I}) has been solved by the truncated Painlev\'e
approach and the fields $q$ and $r$ can be given in terms
of the arbitrary functions as
\numparts
\begin{eqnarray}
q=\frac{q_0}{\phi_1(x,y)+F_2(y)+F_3(t-y)}, \label{p17} \\
r=\frac{\phi_{1x}\phi_{2t}}{q_0 (\phi_1(x,y)+F_2(y)+F_3(t-y))} \label{p18}
\end{eqnarray} 
\endnumparts
and
\begin{equation}
L=\frac{2 \phi_{1x}^2}{(\phi_1(x,y)+F_2(y)+F_3(t-y))^2}-
\frac{2 \phi_{1xx}}{(\phi_1(x,y)+F_2(y)+F_3(t-y))}+L_2,  \label{p19} 
\end{equation}
where
\numparts
\begin{eqnarray}
L_2=
\int\frac{1}{2}\bigg( \frac{i(\phi_{1yy}+F_{2yy})+\phi_{1xxy}}{\phi_{1x}}
-\frac{i(\phi_{1y}+F_{2y})+\phi_{1xx}}{\phi_{1x}^2}\phi_{1xy} \bigg)dx \nonumber\\
\;\;\;\;\;\;\;\;\;\;\; +\frac{1}{2}\frac{i\phi_{1xy}+\phi_{1xxx}}{\phi_{1x}}
-\frac{1}{4}\frac{(\phi_{1y}+F_{2y})^2+\phi_{1xx}^2}{\phi_{1x}^2}.\nonumber
\end{eqnarray} 
\endnumparts
Here the function $\phi_2(y,t)$ is given by the equation (\ref{p16}) and
$q_0$ by (\ref{p12}), while the functions $\phi_1(x,y)$, $F_2(y)$, $F_3(t-y)$ are
themselves arbitrary in the indicated variables.

\section{Novel exact solutions of LSRI equation}
Now we make use of the above truncated Laurent expansion solution to
obtain exact solutions of the LSRI equation (\ref{I}) for the variables
$S(x,y,t)$ and $L(x,y,t)$.  

Taking into account our notation in equation (\ref{s1}), that is $q=S(x,y,t)$ and $r=S^*(x,y,t)$,  
we have $q=r^*$ as far as equation (\ref{I}) is concerned.  Using this condition in equations (\ref{p17}) and 
(\ref{p18}), we obtain the condition, 
\begin{equation}
[F_1(t-y)]^2=F_{3t}.
\end{equation}
Thus, from the results of the previous section, we find that the solution of the
original variable $S(x,y,t)$ takes the form
\begin{equation}
S(x,y,t)=\frac{\sqrt{F_{3t} \phi_{1x}}e^{\int\frac{1}{2}\frac{i(\phi_{1y}+F_{2y})}
{\phi_{1x}}{\rm d}x}}{(\phi_1(x,y)+F_2(y)+F_3(t-y))},  \label{p20}
\end{equation}
while its squared magnitude takes the form
\begin{equation}
|S|^2=\frac{\phi_{1x}F_{3t}}{ (\phi_1(x,y)+F_2(y)+F_3(t-y))^2}. \label{p21}
\end{equation}
The form of $L(x,y,t)$ remains the same as given in equation (\ref{p19}).
With the above general form of the solutions, we now identify interesting classes of
exact solutions to equation (\ref{I}), including periodic and localized solutions by giving
specific forms for the three arbitrary functions $\phi_1(x,y)$, $F_2(y)$ and $F_3(t-y)$.

\subsection{Periodic solutions and localized dromion solutions}

Let us now choose the arbitrary functions $\phi_1$ and $F_3$ to be  Jacobian elliptic
functions, namely $sn$ or $cn$ functions.  The motivation behind this choice of
arbitrary function stems from the fact that the limiting forms of these
functions happen to be localized functions.  Hence, a choice
of $cn$ and $sn$ functions can yield periodic solutions which are more general
than exponentially localized solutions (dromions).  We choose, for example,
\begin{equation}
\phi_1=\mbox{sn}(ax+by+c_1;m_1),\ F_2=4,\ F_3=\mbox{sn}(t-y+c_2;m_2) \label{l1}
\end{equation}
so that
\begin{equation}
S(x,y,t)=\frac{\sqrt{a \mbox{cn}(u_1;m_1) \mbox{dn}(u_1;m_1) 
\mbox{cn}(u_2;m_2) \mbox{dn}(u_2;m_2)}}{(4 + \mbox{sn}(u_2;m_2) +
\mbox{sn}(u_1;m_1))}  e^{\frac{ib}{2a}x}\label{l1_2},
\end{equation}
where $u_1=ax + by+c_1$ and $u_2=t-y+c_2$.
In equations (\ref{l1}) and (\ref{l1_2}), the quantities $m_1$ and $m_2$ are the
modulus parameters of the respective Jacobian elliptic functions while
$a$, $b$, $c_1$ and $c_2$ are arbitrary constants.
The corresponding expression for $|S(x,y,t)|^2$ takes
the form
\begin{equation}
|S|^2=\frac{|a \mbox{cn}(u_1;m_1) \mbox{dn}(u_1;m_1) 
\mbox{cn}(u_2;m_2) \mbox{dn}(u_2;m_2)|}{(4 + \mbox{sn}(u_2;m_2) +
\mbox{sn}(u_1;m_1))^2}. \label{q}
\end{equation}
The profile 
of the above solution for the parametric choice $a$=$b$ =1, 
$c_1$=$c_2$=0,  $m_1$= 0.2, $m_2$ = 0.3, $t$=0 is shown in figure 1(a).
Note that the periodic wave moves with unit phase velocity. 

\subsubsection{(1,1) dromion solution} 
As a limiting case of the periodic solution given by  equation (\ref{q}), 
when $m_1$, $m_2$ $\rightarrow$ 1, the above 
solution degenerates into an exponentially localized solution (dromion).
Noting that $cn(u;1)=dn(u;1)=\mbox{sech}u$ and $sn(u;1)=\mbox{tanh}u$, 
the limiting forms corresponding to (1,1) dromion take the expressions
\begin{equation}
S=\frac{\sqrt{a} \mbox{sech}(t-y+c_2) \mbox{sech}(ax+by+c_1)}
{4+\mbox{tanh}(ax+by+c_1)
+\mbox{tanh}(t-y+c_2)}e^{\frac{ib}{2a}x}
\label{l3}
\end{equation}
and
\begin{equation}
|S|^2=\frac{a \mbox{sech}^2(t-y+c_2) \mbox{sech}^2(ax+by+c_1)}
{(4+\mbox{tanh}(ax+by+c_1)+\mbox{tanh}(t-y+c_2))^2}.
\label{l3_1}
\end{equation}
The variable $L$ then takes the form (using expression (\ref{p19}))
\begin{eqnarray}
L&=&\frac{2a^2 \mbox{sech}^4(ax+by+c_1)}
{(4+\mbox{tanh}(ax+by+c_1)+\mbox{tanh}(t-y+c_2))^2} \nonumber \\
&&-\frac{2\mbox{sech}^2(t-y+c_2)} 
{(4+\mbox{tanh}(ax+by+c_1)+\mbox{tanh}(t-y+c_2))}+ \nonumber \\
&&-\mbox{tanh}(ax+by+c_1)-i \mbox{tanh}(ax+by+c_1)\nonumber \\
&&+\mbox{tanh}^2(ax+by+c_1)-\mbox{sech}^2(ax+by+c_1)-\frac{1}{4} \label{l3_2}
\end{eqnarray}
Schematic form of the (1,1) dromion for the parametric choice $a$=$b$=1, 
$c_1$=$c_2$=0  is shown in figure 1(b). Again note that
the dromion travels with unit velocity in a diagonal direction in the $x-y$ plane.
One can check that the (1,1) dromion obtained by Lai and Chow in reference 8 using the Hirota's bilinear method is a special case of the above solution (\ref{l3_1}) by fixing the parameters $a$, $b$, 
$c_1$ and $c_2$ suitably.  However the later method is unable to give more general solutions (see also Appendix A).

\subsection{More general periodic solutions and higher order dromion solutions}
\subsubsection{Periodic solution and (2,1) dromion} 
Next we obtain more general periodic solution by choosing further general 
forms for the arbitrary functions.  As an example, we choose
\begin{eqnarray}
\phi_1=d_1 \mbox{sn}(c_1+a_1x+b_1y;m_1)+d_2
\mbox{sn}(c_2+a_2x+b_2y;m_2),\nonumber \\
\;\;\;\;\;\; F_2=4,\ F_3=d_3 \mbox{sn}(c_3+t-y;m_3). \label{l4}
\end{eqnarray} 
where $a_i$, $b_i$, $c_i$ and $d_i$ are arbitrary constants and $m_i$'s are modulus parameters ($i=1,2,3$).
Then
\begin{equation}
|S|^2= \frac{q_1}{q_2},  \label{l4_1}
\end{equation}
where $q_1=|(d_1 a_1\mbox{cn}(u_1;m_1)\mbox{dn}(u_1;m_1)+d_2 a_2\mbox{cn}(u_2;m_2)\mbox{dn}(u_2;m_2))
d_3 \mbox{cn}(u_3;m_3)$ $\mbox{dn}(u_3;m_3)|$, $q_2=(4+d_1 \mbox{sn}(u_1;m_1)+d_2
\mbox{sn}(u_2;m_2)+d_3 \mbox{sn}(u_3;m_3))^2$, $u_1=c_1+a_1x+b_1y$, $u_2=c_2+a_2x+b_2y$ and $u_3=c_3+t-y$ with corresponding expressions for $S(x,y,t)$.
The profile of the above solution for the parametric choice $a_1$ =1,
$b_1$ =1, $a_2$ =1,$b_2$ =-1,$d_1$=5, $d_2$=4, $d_3$=0.5, $c_1$=0, $c_2$=$c_3$=5, $m_1$= 0.2, $m_2$ = 0.3, $m_3$ = 0.4, $t$=0 is shown in
figure 2a.  
As  $m_1$, $m_2$, $m_3$ $\rightarrow$ 1, the above solution, namely equation (\ref{l4_1}),
degenerates into a
(2,1) dromion solution given by
\begin{equation}
|S|^2= \frac{(d_1 a_1\mbox{sech}^2u_1+d_2 a_2\mbox{sech}^2u_2)
d_3 \mbox{sech}^2u_3}
{(4+d_1 \mbox{tanh}u_1+d_2
\mbox{tanh}u_2+d_3 \mbox{tanh}u_3)^2} \label{l4_2}
\end{equation}
where $u_1=c_1+a_1x+b_1y$, $u_2=c_2+a_2x+b_2y$ and $u_3=c_3+t-y$.
 The dromion interaction for the parametric choice 
$a_1$=$b_1$=$a_2$ =1, $b_2$ =-1, $d_1$=0.5, $d_2$=$d_3$=1, $c_1$=$c_2$=$c_3$=0
is shown in figures 2b-2d for different time
intervals. Here both the dromions travel with equal velocity but
along opposite diagonals in the $x-y$ plane. The interaction is 
elastic for this choice. The variable $L$ can be evaluated again using equation (31), which we desist
from presenting here.

\subsubsection{Periodic solution and (2,2) dromion} 
Another example for more general periodic solution 
is given by choosing
\begin{eqnarray}
\phi_1=d_1 \mbox{sn}(c_1+a_1x+b_1y;m_1)+d_2
\mbox{sn}(c_2+a_2x+b_2y;m_2),\nonumber \\
\;\;\;\;\;\; F_2=4,\ F_3=d_3 \mbox{sn}(c_3+t-y;m_3)+d_4 \mbox{sn}(c_4+t-y;m_4) \label{l4_3}
\end{eqnarray} 
In equation (\ref{l4_3}),  we choose $m_1$, $m_2$, $m_3$ $\rightarrow$ 1,
to obtain (2,2) dromion solution given by 
\begin{equation}
|S|^2= \frac{(d_1 a_1\mbox{sech}^2u_1+d_2 a_2\mbox{sech}^2u_2)
(d_3 \mbox{sech}^2u_3+d_4 \mbox{sech}^2u_4)}
{(4+d_1 \mbox{tanh}u_1+d_2
\mbox{tanh}u_2+d_3 \mbox{tanh}u_3+d_4 \mbox{tanh}u_4)^2}, \label{l4_4}
\end{equation}
where $u_1=c_1+a_1x+b_1y$, $u_2=c_2+a_2x+b_2y$, $u_3=c_3+t-y$ and $u_4=c_4+t-y$. 
The solution of (2,2) dromion for the parametric choice 
$a_1$=$b_1$=$a_2$ =$b_2$ =1, $d_1$=0.5, $d_2$=$d_3$=$d_4$=1, $c_1$=$c_2$=$c_3$=$c_4$=0
is plotted in figure 3 for various time intervals.  We find that there are
two sets of dromions, each set
containing two dromions one followed by the other. The two sets of dromions are travelling
with same velocity in opposite diagonals of the x-y plane. The dromions 
interact and move forward as time progresses.

\subsubsection{(M,N) dromion}
To generalize the above solutions, we choose
\numparts
\begin{eqnarray}
\phi_1=\sum_{j=1}^M d_j \mbox{sn}(c_j+a_jx+b_jy;m_j), \\
F_2=4,  F_3=\sum_{k=1}^N d_k \mbox{sn}(c_k+t-y;m_k) 
\end{eqnarray} 
\endnumparts
where $a_j$, $b_j$, $c_j$, $d_j$, $c_k$, $d_k$ are arbitrary constants and all
$m_j$'s and $m_k$'s take values between 0 to 1 for periodic solutions and
equal to 1 for dromion solutions.
We proceed as above to construct 
periodic and dromion solutions respectively as
\begin{equation}
|S|^2=\frac{|\sum_{j=1}^M d_ja_j \mbox{cn}(u_1;m_j)\mbox{dn}(u_1;m_j)
\sum_{k=1}^N d_k \mbox{cn}(u_2;m_k)\mbox{dn}(u_2;m_k)|}
{4+\sum_{j=1}^M d_j \mbox{sn}(u_1;m_j)+\sum_{k=1}^N d_k
\mbox{sn}(u_2;m_k)}
\end{equation}
and 
\begin{equation}
|S|^2=\frac{\sum_{j=1}^M d_ja_j \mbox{sech}^2(c_j+a_jx+b_jy)
\sum_{k=1}^N d_k \mbox{sech}^2(c_k+t-y)}
{4+\sum_{j=1}^M d_j \mbox{tanh}(c_j+a_jx+b_jy)+\sum_{k=1}^N d_k
\mbox{tanh}(c_k+t-y)}
\end{equation}
where $u_1=c_j+a_jx+b_jy$ and $u_2=c_k+t-y$.

\subsection{Instanton type solutions}
Another type of elliptic function solution can be chosen as 
\begin{equation}
\phi_1=\mbox{sn}(ax+c_1;m_1)\mbox{cn}(by+c_2;m_2),\ F_2=4,\ F_3=\mbox{sn}(t-y+c_3;m_3).
\label{l2}
\end{equation}
Then
\begin{equation}
|S|^2=\frac{|a\mbox{cn}(u_1;m_1)\mbox{dn}(u_1;m_1)\mbox{cn}(u_2;m_2)
\mbox{cn}(u_3;m_3)\mbox{dn}(u_3;m_3)|}
{(4+\mbox{sn}(u_1;m_1)\mbox{cn}(u_2;m_2)+\mbox{sn}(u_3;m_3))^2}. \label{l2_1}
\end{equation}
where $u_1=ax+c_1$, $u_2=by+c_2$ and $u_3=t-y+c_3$.
The profile of the above periodic solution for the parametric choices 
$a$=1,$b$ =-1, $c_1$=$c_2$=$c_3$=0
$m_1$ = 0.2, $m_2$=0.3, $m_3$=0.4 is 
shown in figure 4a. \\

As $m_1, m_2, m_3 \rightarrow 1$, equation (\ref{l2_1}) degenerates into
an instanton type solution, 
\begin{equation}
|S|^2=\frac{a\mbox{sech}^2(t-y+c_3)\mbox{sech}^2(ax+c_1)\mbox{sech}(by+c_2)}
{(4+\mbox{tanh}(ax+c_1)\mbox{sech}(by+c_2)+\mbox{tanh}(t-y+c_3))^2}. \label{l2_2}
\end{equation}
Schematic diagram of the instanton solution for the parametric choice 
$a$=1,$b$ =-1, $c_1$=$c_2$=0,$c_3$=0.5 is shown in figures 4b-4f for various
time intervals.
We can see that the instanton expressed by (\ref{l2_2}),
has maximum amplitude at $t=0$ while the amplitude decays exponentially
as time $|t| \rightarrow \infty$.

\subsection{2-instanton solution}
A more general form of (\ref{l2_1}) is given by
\numparts
\begin{eqnarray}
\phi_1=d_1 \mbox{sn}(c_1+a_1x;m_1)\mbox{cn}(b_1y;m_2)+d_2
\mbox{sn}(c_2+a_2x;m_3)\mbox{cn}(b_2y;m_4), \\
F_2=4, F_3=d_3 \mbox{sn}(c_3+t-y;m_5). 
\end{eqnarray} 
\endnumparts
Then
\begin{equation}
|S|^2=\frac{f_1}{f_2}. \label{l2_3}
\end{equation}
Here $f_1=|(d_1a_1\mbox{cn}(u_1;m_1)\mbox{dn}(u_1;m_1)\mbox{cn}(b_1y;m_2)+
d_1a_2\mbox{cn}(u_2;m_3)\mbox{dn}(u_2;m_3)\mbox{cn}(b_2y;m_4))$
$d_3 \mbox{cn}(u_3;m_5)\mbox{dn}(u_3;m_5)|,
f_2=(4+d_1 \mbox{sn}(u_1;m_1)\mbox{cn}(b_1y;m_2)+d_2\mbox{sn}$
$(u_2;m_3)\mbox{cn}(b_2y;m_4)+d_3 \mbox{sn}(u_3;m_5))^2,
u_1=c_1+a_1x, u_2=c_2+a_2x$, and $u_3=c_3+t-y$.
The above periodic solution for the parametric choices 
$a_1$=$b_1$=$a_2$=1, $b_2$ =-1, $d_1$=$d_2$=$d_3$=1, $c_1$=$c_3$=-5,$c_2$=0,
$m_1$ = 0.2, $m_2$=0.3, $m_3$=0.4, $m_4$=0.2, $m_5$=0.3 is 
shown in figure 5a. 

As $m_1, m_2, m_3, m_4, m_5 \rightarrow 1$, equation (\ref{l2_3}) degenerates into
two instanton solution given by
\begin{equation}
|S|^2=\frac{(\mbox{sech}^2(a_1x)\mbox{sech}(b_1y)+\mbox{sech}^2(a_2x)\mbox{sech}(b_2y))\mbox{sech}^2(t-y)}
{(4+\mbox{tanh}(a_1x)\mbox{sech}(b_1y)+\mbox{tanh}(a_2x)\mbox{sech}(b_2y)+\mbox{tanh}(t-y))^2}. 
\label{l2_4}
\end{equation}
The time evolution of the solution (\ref{l2_4}) is shown in figures 5b-5f. 
Choosing the arbitrary constants appropriately, we have
one of the instanton having a maximum amplitude at t=-2 while
the other at t=2 and decay exponentially as time $|t| \rightarrow \infty$.

To generalize the above solutions, we choose
\numparts
\begin{eqnarray}
\phi_1=\sum_{j=1}^M d_j \mbox{sn}(c_j+a_j x;m_j)\mbox{cn}(f_j+b_j y;n_j), \\
F_2=4,  F_3=\sum_{k=1}^N d_k \mbox{sn}(c_k+t-y;m_k) 
\end{eqnarray} 
\endnumparts
where $a_j$, $b_j$, $c_j$, $d_j$, $f_j$, $c_k$, $d_k$ are arbitrary constants,
$m_j$, $n_j$ and $m_k$ take values between 0 and 1.
One can construct multi-instanton solution by choosing all the values of
$m_j$, $n_j$ and $m_k$ to be equal to 1. 

\subsection{Bounded multiple solitary waves} 
In the expression (\ref{p21}), one can also easily identify bounded multiple solitary waves all moving with the same velocity.  For instance, using the Jacobian elliptic function form (\ref{l4_3}) with $d_2=0$ in the limit $m_1, m_3, m_4 \rightarrow 1$, one can obtain multiple solitary waves which are bounded.  Figure 6 displays the
structure of a two-soliton solution expressed by
\begin{equation}
|S|^2=\frac13\frac{{\rm sech}^2\frac{x+y}{3}[{\rm
sech}^2(t-y+5)+2{\rm
sech}^2(t-y-5)]}{[\tanh\frac{x+y}3+8+\tanh(t-y+5)+2\tanh(t-y-5)]^2}
\end{equation}
which corresponds to the selections (see equation (\ref{l4_3}))
\begin{equation}
\phi_1=\frac12\tanh\frac{x+y}3,\ F_2=4,\
F_3=\frac12\tanh(t-y+5)+\tanh(t-y-5) \label{l5}
\end{equation}
The figure shows that one of the solitary wave follows the other one but both
are travelling with equal velocity. Hence, there will not be any interaction
between them. 

Finally, one can obtain other interesting classes of solutions for different choices of the arbitrary functions in equations (\ref{p20}), (\ref{p21}) and (\ref{p19}).

\section{Conclusion}
In summary, we have investigated the singularity structure of the (2+1) 
dimensional LSRI equation and confirmed that it satisfies the Painlev\'e property.  The
Painlev\'e truncation approach has been used to construct successfully a very wide class
of solutions of the (2+1) dimensional LSRI equation. The rich solution
structure of the LSRI equation is revealed because of the entrance
of three arbitrary functions in (34) and (31). Especially, Jacobian elliptic
function periodic wave solutions and three special
localized structures, namely dromion, dromion type instanton and 
bounded dromion solutions are given explicitly. However,
more general multiple non-bounded dromion solutions whose phase
velocities differ from each other have not yet been obtained from
the present approach.  It appears that one has to solve equation (22) for more general solutions than the form (23) presented in this paper in order to deduce more general solution.  This is an open problem at present. 

\ack
The work of C. S. and M. L. form part of a Department of Science and
Technology, Govt. of India sponsored research project.
The work of S. Y. L. was supported by the National Natural Science Foundations
of China (No. 90203001 and 10475077). R. R. wishes to thank Department of Science and Technology (DST). 
Govt. of India for sponsoring a major research project.

\begin{appendix}
\section{One dromion solution through Hirota Bilinearization}
Here we briefly point out how the (1,1) dromion solution can be obtained through Hirota bilinearization
method [8].
To bilinearize equation (1), we make the transformation
$$
S= \frac{g}{f}, \;\;\; L=2(\mbox{log}f)_{xx}, \eqno{(A.1)}
$$
which can be identified from the Painlev\'e analysis in Sec. 2.  The resultant bilinear form is
given by
$$
(i(D_t+D_y)+D_x^2)g . f =0, \eqno{(A.2a)}
$$
$$
D_xD_t f . f =2 g g^*  \eqno{(A.2b)}
\label{A1}
$$
where $D$'s are the usual Hirota operators.
To generate a (1,1) dromion, one considers the ansatz
$$
f = 1+e^{\psi_{1}+\psi_{1}^*}+e^{\psi_{2}+\psi_{2}^*}
+M e^{\psi_{1}+\psi_{1}^*+\psi_{2}+\psi_{2}^*}, \eqno{(A.3)}
$$
where
$$
\psi_{1} = px+qy, \eqno{(A.4a)}
$$
$$
\psi_{2} = \lambda y-\Omega t. \eqno{(A.4b)}
$$
Here $M$ is a real constant and $p, \Omega, \lambda$ and $q$ are 
complex constants. Substituting (A.3) in (A.2), we obtain
$$
g = \rho e^{\psi_{1}+\psi_{2}}, \eqno{(A.5a)}
$$
$$
|\rho|^2  =  (p+p^*)(q+q^*)(1-M) \eqno{(A.5b)}
$$
and also the conditions $M<1$, $q=ip^2$ and $\Omega=\lambda$.
This is a special case of the dromion we have obtained in (\ref{l3_1})
with the constants $c_1=c_2=\frac{1}{2}\mbox{log}\frac{3}{2}$ and
by choosing $\psi_{1R}=-(ax+by+c_1)$ and $\psi_{2R}=-(t-y+c_2)$ and $M=\frac{1}{3}$
where  $\psi_{1R}, \psi_{2R}$ are the real parts of $\psi_{1}, \psi_{2}$ 
respectively. Thus, equation (\ref{l3_1}) contains the solution of Lai
and Chow [8] as a special case.  No higher order solution has been constructed by this method.
\end{appendix}

\vspace{1cm}

{\bf Figure captions} \\
{\bf Figure 1} (a) Elliptic function solution (\ref{q}) (b) Localized dromion solution (\ref{l3_1}) for the variable  $|S(x,y,t)|^2$ (c) the corresponding magnitude of the variable $L(x,y,t)$ given by (\ref{l3_2}) \\
{\bf Figure 2} (a) Elliptic function solution (\ref{l4_1}), (b-d) (2,1) dromion solution (\ref{l4_2}) and its interaction at time intervals (b) $t=-10$, (c) $t=0$ and (d) $t=10$ \\
{\bf Figure 3} (a-e) (2,2) dromion solution (\ref{l4_4})  interaction at time intervals
(a) $t=-8$, (b) $t=-4$ (c) $t=0$ (d) $t=4$  and (e)$t=8$ \\
{\bf Figure 4} (a) Elliptic function solution (\ref{l2_1}) and an instanton solution (\ref{l2_2}) at time intervals (b) $t=-3$, (c) $t=-1$ and (d) $t=1$  (e) $t=3$ and (f) $t=4$ \\
{\bf Figure 5} (a) Elliptic function solution (\ref{l2_3}) and two instanton solution (\ref{l2_4}) at time 
intervals (b) $t=-3$, (c) $t=-1$, (d) $t=1$,  (e) $t=3$ and (f) $t=4$  \\
{\bf Figure 6} Bounded two soliton solution


\Bibliography{15}
\bibitem{ref1}
Boiti M, Leon J J P,  Martina L  and Pempinelli F, 1988 {\it Phys. Lett. A} {\bf 132} 432  
\bibitem{ref2}
Fokas A S and Santini P M, 1990 {\it Physica D} {\bf 44} 99  
\bibitem{ref3}
Konopelchenko B G 1993 {\it Solitons in Multidimensions} (Singapore: World Scientific)
\bibitem{ref4}
Radha R and Lakshmanan M 1994 {\it J. Math. Phys.} {\bf 35} 4746 
\bibitem{ref5}
Radha R and Lakshmanan M 1995 {\it J. Phys. A: Math. Gen.} {\bf 28} 6977 
\bibitem{ref6}
Radha R and Lakshmanan M 1996 {\it J. Phys.  A: Math. Gen.} {\bf 29} 1551 
\bibitem{ref7_1}
Oikawa M, Okamura M and Funakoshi M 1989 {\it J. Phys. Soc. Japan} {\bf 58} 4416
\bibitem{ref7}
Lai D W C and Chow K W 1999 {\it J. Phys. Soc. Japan} {\bf 68} 1847
\bibitem{ref8}
Lou S Y 1995 {\it J. Phys. A: Math. Gen.} {\bf 28} 7227 
\bibitem{ref9}
Lou S Y 2000 {\it Phys. Lett. A} {\bf 277} 94 
\bibitem{ref10}
Lou S Y and Ruan H Y 2001 {\it J. Phys. A: Math. Gen.} {\bf 34} 305 
\bibitem{ref11}
Tang X Y, Lou S Y and Zhang Y 2002 {\it Phys. Rev. E.} {\bf 66} 046601
\bibitem{ref12}
Lou S Y, Chen C L and Tang X Y 2002 {\it J. Math. Phys.}  {\bf 43} 4078
\bibitem{ref13}
Tang X Y and Lou S Y 2003 {\it J. Math. Phys.} {\bf 44} 4000
\endbib
\end{document}